%% file: accelerated_torque_calculation.tex
\documentclass[journal,transmag]{IEEEtran}
\usepackage{hyperref,color,multirow}
\usepackage{geometry}
\usepackage{graphicx,color,epsf}
\usepackage{amsmath,amsfonts,amssymb,amscd,bm}
\usepackage{algorithm}
\usepackage{algpseudocode}
\usepackage{subcaption}
\usepackage[english]{babel}
\usepackage[utf8]{inputenc}
\usepackage{booktabs}
\usepackage[normalem]{ulem}
\makeatletter
\newcommand{\removelatexerror}{\let\@latex@error\@gobble}
\makeatother

\usepackage{eso-pic}

\begin{document}
	\newgeometry{top=0.7in, left=0.65in, textheight=9.2in, textwidth=7.2in}
	
	\title{Accelerated {Steady-State} Torque Computation for Induction Machines using Parallel-In-Time Algorithms}
	
	\author{\IEEEauthorblockN{
			Denys Bast\IEEEauthorrefmark{1},
			Iryna Kulchytska-Ruchka\IEEEauthorrefmark{1},
			Sebastian Sch\"{o}ps\IEEEauthorrefmark{1}, and
			Oliver Rain\IEEEauthorrefmark{2}}
		
		\IEEEauthorblockA{
			\IEEEauthorrefmark{1}Technische Universit\"{a}t Darmstadt, Institut f\"{u}r Teilchenbeschleunigung und Elektromagnetische Felder,\\Schlossgartenstrasse 8, D-64289 Darmstadt, Germany
		}
		
		\IEEEauthorblockA{
			\IEEEauthorrefmark{2}Robert Bosch GmbH,\\Robert-Bosch-Campus 1, D-71272 Renningen, Germany
		}

	}%
	
	\IEEEtitleabstractindextext{%
		\begin{abstract} 
			This paper focuses on efficient steady-state computations of induction machines. In particular, the periodic Parareal algorithm with initial-value coarse problem (PP-IC) 
			is considered for acceleration of classical time-stepping simulations via non-intrusive parallelization in time domain, i.e., existing implementations can be reused. 
			Superiority of this parallel-in-time method is in its direct applicability 
			to time-periodic problems, compared to, e.g, the standard Parareal method, which only solves an initial-value problem, starting from a prescribed initial value.
			PP-IC is exploited here to obtain the steady state of several operating points of an induction motor, developed by Robert Bosch GmbH. Numerical experiments show that acceleration up 
			to several dozens of times can be obtained, depending on availability of parallel processing units. Comparison of PP-IC with existing time-periodic explicit error correction method
			highlights better robustness and efficiency of the considered time-parallel approach.
		\end{abstract}
		
		\begin{IEEEkeywords}
			Parallel-in-time, {steady state}, eddy currents, time stepping
	\end{IEEEkeywords}}

	\maketitle
	\thispagestyle{empty}
	\pagestyle{empty}
	\section{Introduction}
\begin{figure}[t]
		\centering
		\includegraphics[width=.39\textwidth]{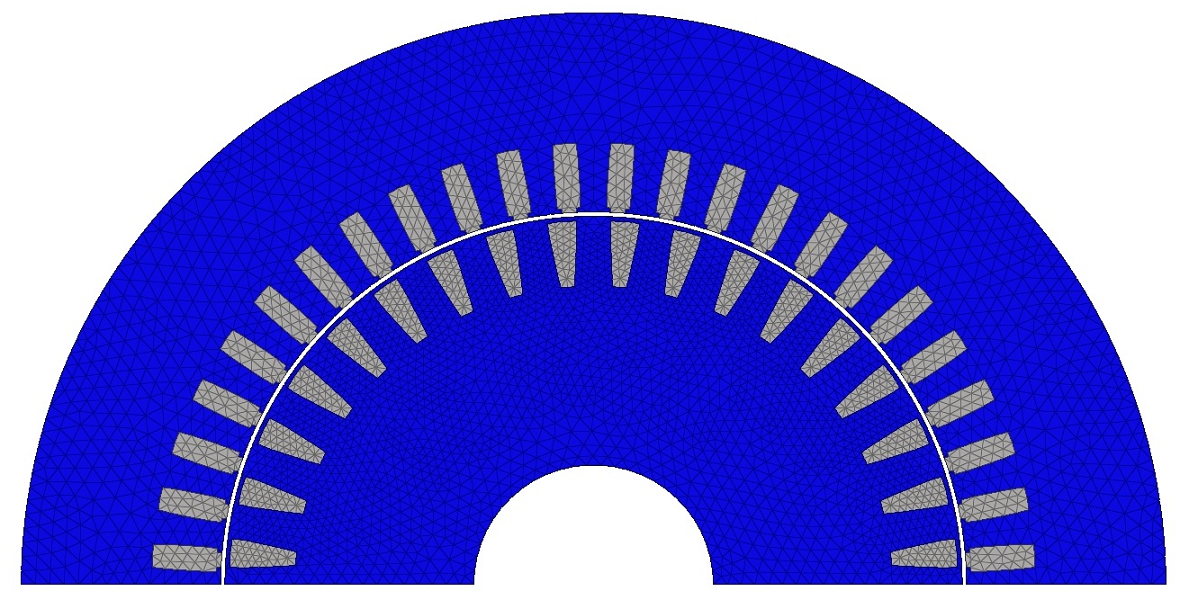}
		\caption{Finite element model of an induction motor of electric vehicle drive, provided by Robert Bosch GmbH.}
		\label{fig:asmBsp}
	\end{figure}

  Induction machines are used in a wide variety of industrial applications. {Their operation covers the} power ranges from hundreds of watts to several megawatts. For the design of these motors, engineers are often interested in the steady-state operating characteristics {such as, e.g.,} mean torque, efficiency, periodically changing currents and voltages at certain revolution speeds.
	{During initial design stages, information about these quantities can be obtained from numerical analysis of transient eddy current problems} using space and time discretization {\cite{Arkkio_1990aa}}, see, e.g., Figure~\ref{fig:asmBsp}. {Application} of an implicit time stepping leads to a nonlinear system of equations at each time step. Numerical simulation of induction motors is often computationally expensive because a lot of time steps {might} need to be calculated to reach the steady state. Therefore, {efficient numerical} methods {are necessary} {to accelerate} these calculations. 

{Different methodologies have been developed in order to achieve the periodic steady-state solution. For instance, the time-periodic finite element method \cite{Nakata_1995aa} requires solution of a coupled periodic system, which could be prohibitively expensive for modern real-life applications. Accelereted convergence to the steady state with the simplified \textit{time-periodic explicit error correction method (TP-EEC)} \cite{Takahashi_2010aa, Katagiri_2011aa} is based on correction of the sequential solution after each (half-) period. However, it performs well only for a specific class of problems, particularly when the time constant is big enough.}
	
	The Parareal algorithm \cite{Lions_2001aa, Gander_2008aa} {is a powerful} approach, {which allows} to speed up the sequential {solution of} the {underlying evolution} problem {via} parallelization in {the time domain.} {The method} has been recently applied to simulation of an {induction} machine by the authors in \cite{Schops_2018aa}, where quick convergence and efficiency of Parareal is illustrated. %
	The \emph{periodic Parareal algorithm with initial-value coarse problem (PP-IC)}{, introduced in} \cite{Gander_2013ab}, is {a natural extension} of the {original} Parareal {to the class of time-periodic problems}. {A multirate version of PP-IC was introduced and applied to a simplified electrical machine model, operated at synchronous speed, by the authors in \cite{Gander_2018aa}. In this paper we investigate the industrial use case of an asynchronous (induction) machine and apply PP-IC to the underlying time-periodic problem. }
	{Performance of the method as well as its comparison with the simplified TP-EEC method is illustrated via steady-state analysis of an induction machine model, developed at the Robert Bosch GmbH, Germany.}

	\section{{Problem Setting} and Discretization}\label{sec:model} %
\begin{figure}[t]
		\centering
		\includegraphics[width=.59\linewidth]{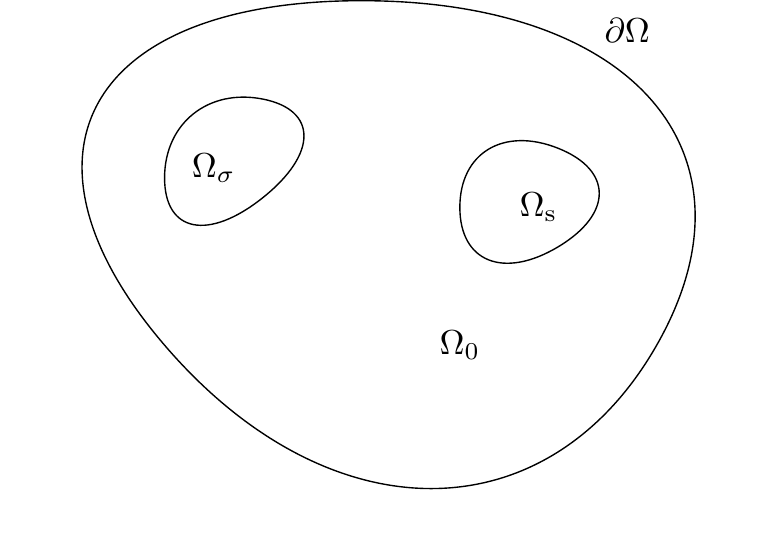}
		\vspace{-1em}
		\caption{Computational domain $\Omega$ of the eddy current problem.}
		\label{fig:pics_patata}
\end{figure}

	The eddy current problem in A$^\star$-formulation with magnetic vector potential $\vec{A}$ is given by 
	\begin{align}
	\sigma\partial_t{\vec{A}}
	+
	\nabla \times (\nu\nabla\times\vec{A})
	=
	\vec{J}_\mathrm{s}(t)
	\label{eq:mqs1}
	\end{align}
	{on} $\Omega\times\mathcal{I}$ with computational domain {$\bar\Omega=\bar\Omega_0\cup\bar\Omega_\sigma\cup\bar\Omega_\textrm{s}$,} depicted in Figure~\ref{fig:pics_patata}, {and time interval} $\mathcal{I}:=(t_0,t_\text{end}].$ 
	{The regions $\Omega_\sigma=\mathrm{supp}(\sigma)$ and $\Omega_\mathrm{s}=\mathrm{supp}(\vec{J}_\mathrm{s})$ denote the conductive ($\sigma>0$) and stranded conductors subdomains. Input currents $i_k$ are homogeneously distributed by means of the stranded-conductor winding functions $\vec{\chi}_{\mathrm{s},k}$ \cite{Schops_2013aa} {and form} the current density $\vec{J}_{\mathrm{s}} = \sum_k\vec{\chi}_{\mathrm{s},k}\,i_k$.} 
	
	{Nonlinearity of the eddy current equation is given by the reluctivity function $\nu=\nu(|\nabla\times\vec{A}|)$. 
	To obtain a well-posed problem we complete \eqref{eq:mqs1} with gauging (e.g., the Coulomb gauge \cite{Monk_2003aa})} 
	and a suitable boundary condition {such as} Dirichlet
		$\vec{n}\times\vec{A}|_\Gamma=0$, where $\Gamma=\partial\Omega$.
	{We also prescribe} an initial value $\vec{A}(\vec{r},t_0)=\vec{A}_0(\vec{r}),$ $\vec{r}\in\Omega$.
	
	\subsection{Computation of the Torque}
	The torque-speed characteristic is one of the most important technical features of an electrical machine. The torque can be calculated with the eggshell method \cite{Henrotte_2004aa}, \cite{Henrotte_2010aa}. {Within this approach} the moving {rigid} piece is surrounded by {a hull $S$, whose thickness does not need to be constant. Movement is then described by the deformation of the eggshell region only.} Using the Maxwell stress tensor $\sigma_\mathrm{EM}$ and the velocity {field $\mathbf{v}$ associated to a displacement of the moving body} by an infinitesimal distance $\delta \mathbf{u},$ {one can write}
	\begin{equation}
	\dot W_\mathrm{EM}= - \mathbf{F} \cdot \delta \dot{\mathbf{u}} =\int_{S} \sigma_\mathrm{EM} :\nabla \mathbf{v} \, \mathrm{d} S
	\end{equation}
{for the mechanical power $\dot W_\mathrm{EM}$ in the eggshell $S$. The velocity and its gradient are determined by} $\mathbf{v}=\gamma \delta \dot{\mathbf{u}}$ {and} $ \nabla \mathbf{v}=\nabla \gamma \delta \dot{\mathbf{u}},$ {respectively, where} $\gamma$ is any smooth function that is equal to $1$ on the inner {surface of $S$} and is $0$ on {its} outer surface. {With these definitions} the resultant force $\mathbf{F}$ {on the moving piece is given by}	
	\begin{equation}
	\mathbf{F}=-\int_{S}\sigma_\mathrm{EM}\cdot \nabla \gamma \mathrm{d}S.
	\label{eggshellFormula}
	\end{equation}
	{The eggshell formula \eqref{eggshellFormula}} can be then used for calculation of the resultant electromagnetic torque $\mathbf{T}$ as
	\begin{equation}
	\mathbf{T}=-\int_{S} \mathbf{r} \times (\nabla \gamma \cdot \sigma_\mathrm{EM})\mathrm{d}S.
	\end{equation} 
	
		\subsection{{Spatial Discretization}}
	{The Ritz-Galerkin approach leads to the following weak formulation for} $\vec{A}\in H_0(\mathrm{curl},\Omega)$
	\begin{align*}
	\int_\Omega \vec{w}\cdot\sigma\partial_t{\vec{A}}
	+
	\nabla\times\vec{w}\cdot(\nu\nabla\times\vec{A})\;\mathrm{d}\Omega
	&=
	\int_\Omega \vec{w}\cdot \vec{J}_{\mathrm{s}}\;\mathrm{d}\Omega
	\end{align*}
	\noindent for all $\vec{w}\in H_0(\mathrm{curl},\Omega)$. 	
  {For a rotating machine, approximation by edge elements \cite{Monk_2003aa} }
	\begin{align*}
	\vec{A}(\vec{x},t)\approx\sum_{i=1}^{n}\vec{w}_i(\vec{x})\;a_i(t)
	\end{align*}
	{yields} the following system of differential algebraic equations (DAEs):
	\begin{align}\label{eq:daes1}
	\mathbf{M}_\sigma\mathrm{d}_t \mathbf{a}(t)
	+
	\mathbf{k}_{\nu}\left(\mathbf{a}(t),\theta(t)\right)
	&=
	\mathbf{j}_{\mathrm{s}}(t)
	\end{align}
	{ for the unknown (line-integrated) magnetic vector potentials $\mathbf{a}(t)\in\mathbb{R}^{n}.$ 
	Here $\mathbf{M}_\sigma\in\mathbb{R}^{n\times n}$ denotes the (singular) mass matrix, function $\mathbf{k}_{\nu}$
	is given by $\mathbf{k}_{\nu}\left(\mathbf{a},\theta\right)=\mathbf{K}_{\nu}\left(\mathbf{a},\theta\right)\mathbf{a}$
	with}	the curl-curl matrix $\mathbf{K}_\nu(\mathbf{a},\theta)\in\mathbb{R}^{n\times n}$, which depends on the rotor angle 
	$\theta,$ and $\mathbf{j}_{\mathrm{s}}(t)\in\mathbb{R}^{n}$ is the discretized source current density.
	
	Rotation is modeled with the moving band approach \cite{Davat_1985aa, Ferreira-da-Luz_2002aa} and is 
	described by the mechanical equations for the angle $\theta$ and angular velocity $\omega$
	\begin{align}
	\label{eq:motion}
	\mathrm{d}_t\theta(t) = \omega(t)
	\quad\text{and}\quad
	I\mathrm{d}_t\omega(t)
	+ {C \omega(t)}
	=
	T\left(\mathbf{a}(t)\right),
	\end{align}
	where $I$ is the {moment of} inertia, {$C$ denotes the friction coefficient,} 
	{and $T$ is the mechanical excitation determined by the magnetic field.}  
	
	{Assigning initial conditions $\theta(t_0)=\theta_0$ and $\omega(t_0)=\omega_0$ 
	together with $\mathbf{a}(0)=\mathbf{a}_0$ and combining \eqref{eq:daes1} and 
	\eqref{eq:motion} we obtain the coupled problem }
	\begin{align}\label{eq:daes3}
	\mathbf{M}\mathrm{d}_t\mathbf{u}(t)
	+
	\mathbf{K}\left(\mathbf{u}(t)\right)\mathbf{u}(t)
	&=
	\mathbf{f}(t)
	\end{align}
	{with respect to the $m:=(n+2)$-dimensional} unknown $\mathbf{u}=\left[\mathbf{a}^{{\!\top}}, \theta, \omega\right]^{\!\top}$. 
	{The matrices $\mathbf{M}$ and $\mathbf{K}$ are given by}
	\begin{equation*}  
	\mathbf{M}=
	\begin{pmatrix}
	\mathbf{M}_{\sigma} & 0 & 0 \\
	{0} & 1 & 0 \\
	{0} & 0 & I
	\end{pmatrix}, \,\,\,\,\,
	\mathbf{K}=
	\begin{pmatrix}
	\mathbf{k}_{\nu}(\cdot,{\theta}) & 0 & 0 \\
	{0} & 0 & -1 \\
	{T(\cdot)} & {0} & {C}
	\end{pmatrix}%
	\end{equation*}
	and the right{-}hand side $
	\mathbf{f}= 
	\left[	\mathbf{j_s}^{\!\top}, 0, {0}
	\right]^{\!\top}.%
	$	
	
	\subsection{Time-Integration Scheme}
	{Since equation \eqref{eq:daes3} is an index-1 DAE \cite{Nicolet_1996aa}, by means of the implicit Euler method it can {essentially} be treated as an ordinary differential equation \cite{Schops_2018aa}.
In this case the time-stepping scheme propagating the solution} from $t_i$ to $t_{i+1} = t_i + \delta t$ {is written as}
	\begin{equation}
	\left(\frac{1}{\delta t} \mathbf{M}+\mathbf{K}\left(\mathbf{u}_{i+1}\right)\right)\, \mathbf{u}_{i+1}=\mathbf{f}_{i+1}+ \frac{1}{\delta t} \mathbf{M} \mathbf{u}_i.
	\label{imEuler}
	\end{equation}
	{This time-integration formula defines a numerical solution operator}
	\begin{equation}
	\label{prop:fine}
	\mathcal{F}: \mathcal{I}\times\mathcal{I}\times \mathbb{R}^{{m}} \rightarrow \mathbb{R}^{{m}}
	\end{equation}
	{such that $\mathbf{u}_{i+1}= \mathcal{F}\left(t_{i+1},t_i,\mathbf{u}_i\right)$.}
	{Similarly, we introduce a coarse propagator}
	\begin{equation}
	\label{prop:coarse}
	\mathcal{G}: \mathcal{I}\times\mathcal{I}\times \mathbb{R}^{{m}} \rightarrow \mathbb{R}^{{m}},
	\end{equation}
	{which, as $\mathcal{F},$ solves the initial-value problem (IVP) for \eqref{eq:daes3}, however uses a lower precision, i.e., a time step $\Delta t \gg \delta t.$}
	%
	\section{Parallel-in-Time Methods}\label{sec:parareal}
	%
\algblock{ParFor}{EndParFor}
\algnewcommand\algorithmicparfor{\textbf{parfor}}
\algnewcommand\algorithmicpardo{\textbf{do}}
\algnewcommand\algorithmicendparfor{\textbf{end\ parfor}}
\algrenewtext{ParFor}[1]{\algorithmicparfor\ #1\ \algorithmicpardo}
\algrenewtext{EndParFor}{\algorithmicendparfor}

\begin{algorithm}[t]
	\caption{The PP-IC algorithm, based on \cite{Gander_2013ab}.}
	\label{alg:ppic}
	\begin{algorithmic}[1]
		
		\State initialize: {$\mathbf{U}_N^{(0)}$} and $\overline{\mathbf{u}}_j^{(0)}, \tilde{\mathbf{u}}_j^{(0)}\leftarrow \mathbf{0}$ (for all $j$);
		\State set counter: $k\leftarrow 1$;
		\While{$k \le {1}\, \mathbf{or}\, \mathrm{max}_j \big|\big|\mathbf{U}_{j}^{(k)}-\mathbf{U}_{j}^{(k-1)}\big|\big| > tol$}
		\State {update: $\mathbf{U}_0^{(k)} \leftarrow \mathbf{U}_N^{(k-1)};$}
		\For{$j \leftarrow 1 ,\, N$}
		\State solve coarse problem: $\overline{\mathbf{u}}_j^{(k)}\leftarrow \mathcal{G}(T_j, T_{j-1}, \mathbf{U}_{j-1}^{(k)})$;
		\State post process: $\mathbf{U}_{j}^{(k)} \leftarrow \tilde{\mathbf{u}}_j^{(k-1)}+\overline{\mathbf{u}}_j^{(k)}-\overline{\mathbf{u}}_j^{(k-1)}$;
		\EndFor
		\ParFor{$j \leftarrow 1 ,\, N$}
		\State solve fine problem: $\tilde{\mathbf{u}}_j^{(k)}\leftarrow \mathcal{F}(T_j, T_{j-1}, \mathbf{U}_{j-1}^{(k)})$;
		\EndParFor
		\State increment counter: $k \leftarrow k+1$;
		\EndWhile
	\end{algorithmic}
\end{algorithm}	
	\subsection{Parareal Algorithm}
	{The main idea of the Parareal method is to parallelize sequential time stepping by distributing the calculations {among} $N$ available CPUs.} First, the time interval $\mathcal{I}$ is divided into $N$ subintervals $\mathcal{I}_j :=(T_{j-1},T_j]$ with $t_0=T_0<T_1 < ...<T_N= t_{\mathrm{end}}$. 
	Equation (\ref{eq:daes3}) has to be solved on each {sub}interval with initial value $\mathbf{U}_{j-1}:=\mathbf{u}(T_{j-1})$ {to obtain the} final value $\mathbf{U}_j:= \mathbf{u}(T_j),$ $j=1,\dots,N.$ {Applying the propagator $\mathcal{F}$ from \eqref{prop:fine} to each IVP the matching conditions at the synchronization points $T_j,$ $j=1,\dots,N$ are imposed as}
	\begin{equation}
	\mathbf{H}(\mathbf{U}):=
	\begin{cases}
	\mathbf{U}_0 - \mathbf{u}_0  &=0,\\
	\mathbf{U}_1-\mathcal{F}\left(T_1,T_0,\mathbf{U}_0\right)&=0,\\
	&\vdots \\
	{\mathbf{U}_{N}-\mathcal{F}\left(T_{N},T_{N-1},\mathbf{U}_{N-1}\right)}&=0.
	\end{cases}
	\label{matchingProblem}
	\end{equation}
{In fact, equation \eqref{matchingProblem} is a root-finding problem for} the nonlinear operator $\mathbf{H}:\mathbb{R}^{(N+1)m}\to\mathbb{R}^{(N+1)m}$ with respect to its {argument $\mathbf{U}=\left[\mathbf{U}^{\!\top}_0,\hdots,\mathbf{U}^{\!\top}_{N}\right]^{\!\top}.$ Application of the Newton method and a finite difference approximation of the Jacobian \cite{Gander_2007aa, Schops_2018aa} with the coarse solver $\mathcal{G}$ defined in \eqref{prop:coarse} gives the Parareal update formula:}	
	\begin{align}
	{\mathbf{U}_0^{(k)}:=} &\;{\mathbf{u}_0,\label{eq:para_init}}\\
	\mathbf{U}_j^{(k)}:= &\;\mathcal{F}(T_j, T_{j-1},{\mathbf{U}^{(k-1)}_{j-1}})\nonumber\\
	 &+\mathcal{G}(T_j, T_{j-1},{\mathbf{U}^{(k)}_{j-1}})-\mathcal{G}(T_j, T_{j-1},{\mathbf{U}^{(k-1)}_{j-1}})\label{eq:para}
	\end{align}
	{for $j=1,\dots,N$ and $k=1,2,\dots,K.$ The pseudo code for \eqref{eq:para_init}-\eqref{eq:para} and a detailed explanation of the iterative procedure is presented in \cite{Schops_2018aa}.} %
	
	\subsection{{PP-IC Iteration}}
	\begin{figure}
		\centering
		\includegraphics[width=0.8\linewidth]{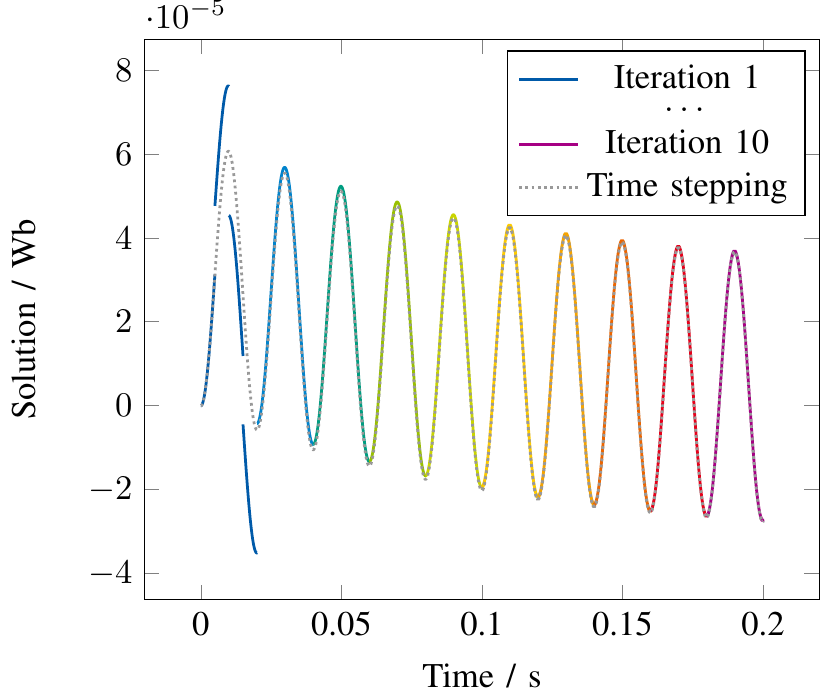}
		\caption{Comparison of PP-IC and time stepping when visualizing the resulting waveform of iteration $k$ on time interval $[t_0+(k-1)T,\;t_0+kT]$ for $k=1,\dots,10$ {with $t_0=0$ and $T=0.02$ s. Illustrated data is based on the solution of a linear RL-circuit model excited with a $50$ Hz sinusoidal current source.}}
		\label{fig:PP-IC_indent}
	\end{figure}%
	{Aiming at} the steady state of an induction machine {we would now like to exploit a version of the Parareal method, adapted to time-periodic problems. Such an algorithm was introduced in \cite{Gander_2013ab} and was called PP-IC.} A time-periodic formulation { for \eqref{eq:daes3} on period $T:=t_\text{end}-t_0$ requires} initial value {$\mathbf{u}(t_0)$} to be {equal to} the final {one $\mathbf{u}(t_\text{end})$. Within the Parareal setting it means substitution of the first equation in \eqref{matchingProblem} for the periodicity condition $\mathbf{U}_0-\mathbf{U}_N=0.$ Analogous derivations as those performed for \eqref{matchingProblem} together with an additional relaxation on the coarse grid give the PP-IC iteration: 
	\begin{align}
	\mathbf{U}_0^{(k)}:= &\;\mathbf{U}_N^{(k-1)},\label{eq:ppic_init}\\
	\mathbf{U}_j^{(k)}:= &\;\mathcal{F}(T_j, T_{j-1},{\mathbf{U}^{(k-1)}_{j-1}})\nonumber\\
	 &+\mathcal{G}(T_j, T_{j-1},{\mathbf{U}^{(k)}_{j-1}})-\mathcal{G}(T_j, T_{j-1},{\mathbf{U}^{(k-1)}_{j-1}})\label{eq:ppic}
	\end{align}
	for $j=1,\dots,N$ and $k=1,2,\dots{,K.}$ It can be seen that PP-IC \eqref{eq:ppic_init}-\eqref{eq:ppic} is based on the Parareal iteration \eqref{eq:para_init}-\eqref{eq:para} with the only difference in the update of the initial value $\mathbf{U}_0^{(k)}$ at iteration $k.$ We note that in contrast to the classical Parareal method, which convergences superlinearly in $K\leq N$ iterations as it is proved in \cite{Gander_2008aa}, convergence of PP-IC is only linear \cite{Gander_2013ab}. }
	
	{The iterative procedure of PP-IC is summarized with the pseudo code in Algorithm \ref{alg:ppic}. Step $4$ updates the solution at the beginning of the period by assigning the end value from the previous iteration. Starting from the corrected $\mathbf{U}_0^{(k)}$ sequential solutions are performed by means of the coarse propagator $\mathcal{G}$ at step $6$ and are expected to be computationally cheap due to a large step size $\Delta t.$ Calculations on the fine grid, obtained via discretization with a small time step $\delta t,$ can be performed for each subinterval in parallel (step $10$), starting from initial values $\mathbf{U}_{j-1}^{(k)},$ $j=1,\dots,N,$ already given from steps $4$ and $7$. We choose both coarse and fine solvers to be the implicit Euler method, using low and high fidelity, respectively.}
			
\begin{figure}[t]
	\centering
	\includegraphics[width=0.8\linewidth]{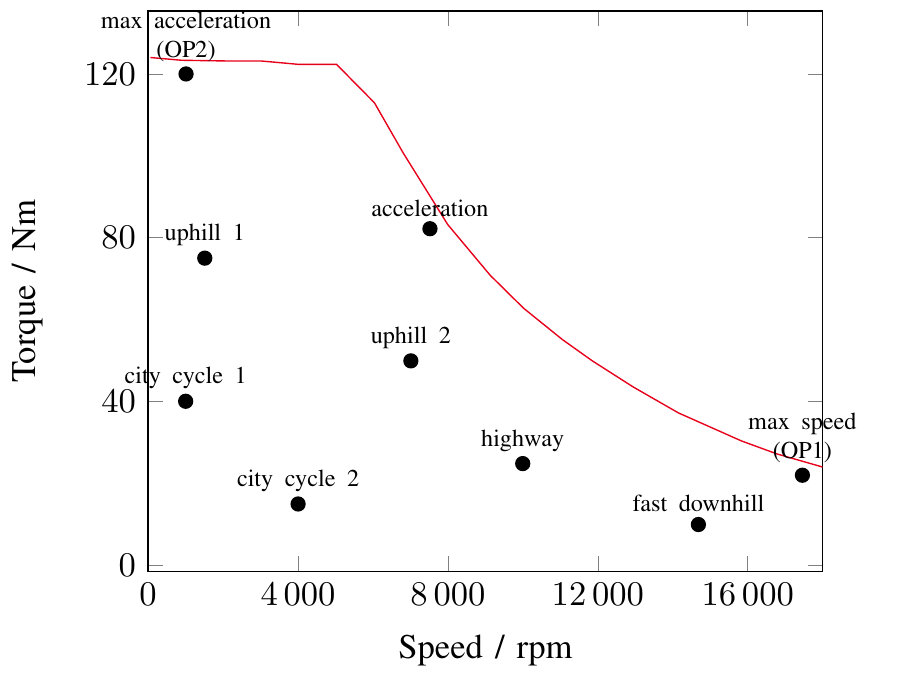}
	\caption{Torque-speed characteristic and some representative operating points of the induction motor from Figure \ref{fig:asmBsp}.}
	\label{fig:torqueSpeed}
\end{figure}%
		
	{PP-IC is an iterative approach applied to a fixed time interval{, on which the periodicity constraint is to be satisfied}. However, PP-IC can be reinterpreted as applying the classical Parareal method on period $[t_0, t_0+T]$ with initial guess $\mathbf{u}_0$ at $t_0$ and using only a single iteration. PP-IC then proceeds to subsequent time intervals $[t_0+(k-1)T, t_0+kT]$ with a single iteration each until the periodicity constraint {together with the matching conditions at every $T_j$ for $j=1,\dots,N-1$ are} fulfilled at some $k=K.$ {We illustrate this interpretation of PP-IC as a forward-in-time Parareal iteration in} Figure~\ref{fig:PP-IC_indent}. {It is visible that the solution after the first iteration of Algorithm~\ref{alg:ppic}, when only initial fine solves have been executed on $N=4$ subintervals in parallel, contains jumps at the synchronization points. However, these discontinuities quickly decrease already after the next PP-IC update at iteration $k=2$ and the obtained solution nearly replicates the classical time stepping solution on the second period $[T, 2T]$ ($T=0.02$ s here). Further PP-IC iterations, in the same way as the standard time-stepping, eventually reduce the periodicity error up to a prescribed tolerance and deliver the periodic steady-state solution.}}

	\subsection{{Estimation of Computational Costs}}
	{For a clear estimation of computational costs of the Parareal algorithms considered in this paper we calculate the number of} effective time steps instead of {the} direct {wall clock} time measurement. {By effective time steps we mean the following.} %
First, assume the number of available processing units is equal to {$N.$} Then we consider the splitting of the time domain into $N$ subintervals with exactly one coarse step per subinterval. Second, parallelization among the $N$ CPUs allows us to take into account fine computations on one subinterval only. We denote the number of fine steps per parallel process by $N_\text{p}$. Effective time steps $N_\text{e}$ after $I_\text{t}$ iterations could then be counted using the formula
	\begin{equation}
	\label{eq:effect_steps}
	N_\text{e} = I_\text{t} \cdot \left(N+N_\text{p}\right).
	\end{equation}
{At every iteration $N+N_\text{p}$} time steps are calculated sequentially. Multiplication with the number of iterations $I_\text{t}$ results in the overall time steps {which} are not executed in parallel.	%

	{We note that the described approach to measure the computational costs is legitimate} {only} if the solution {at} a time step is always {performed} with {a} comparable effort and communication costs are negligible. {In general, this may not hold,} especially since both fine and coarse time steps are considered. Within the {current} implementation {for} the examples presented {in the next section}, the solver always requires $2$-$3$ Newton iterations to solve a time step. Estimate with formula \eqref{eq:effect_steps} is therefore justifiable.
	
	\section{Numerical Results}\label{sec:num}
The torque-speed characteristic of an induction motor is determined by evaluation of some tens to hundreds of operating points, i.e., transient eddy current simulations at a constant revolution speed and mostly sinusoidal coil currents. In electric vehicles such operating points correspond to certain driving conditions. In the following we apply the PP-IC algorithm to the induction motor whose {two-dimensional} mesh view with $4459$ degrees of freedom is depicted in Figure~\ref{fig:asmBsp}. Due to symmetry only half of the geometry {is} modeled and discretized. The torque-speed characteristic as well as some representative operating points of the motor are shown in Figure~\ref{fig:torqueSpeed}.

\begin{figure*}
		\begin{subfigure}{0.5\linewidth}
			\centering
			\includegraphics[width=0.6\linewidth]{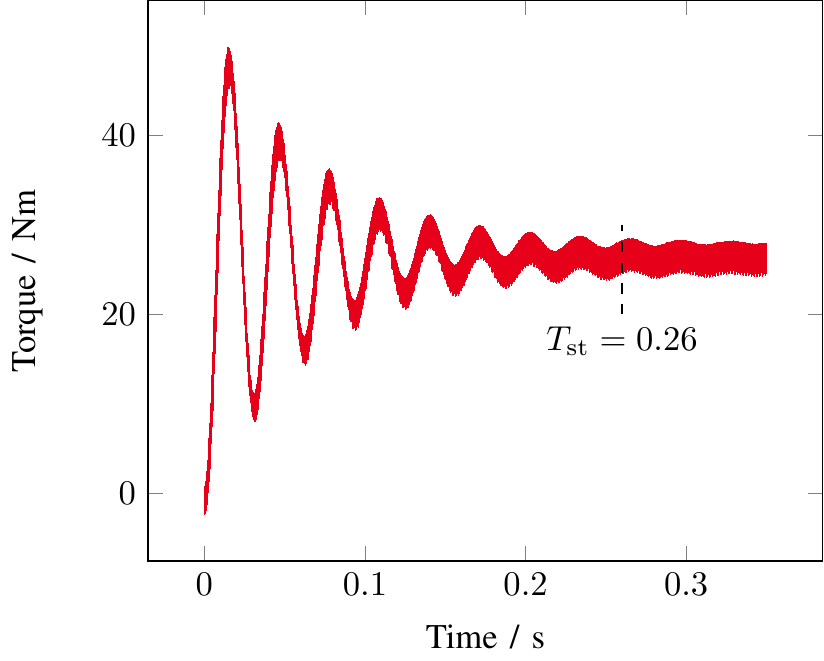}
			\caption{Operating point OP1\label{fig:OP1}}
		\end{subfigure}
		\begin{subfigure}{0.5\linewidth}
			\centering
			\includegraphics[width=0.6\linewidth]{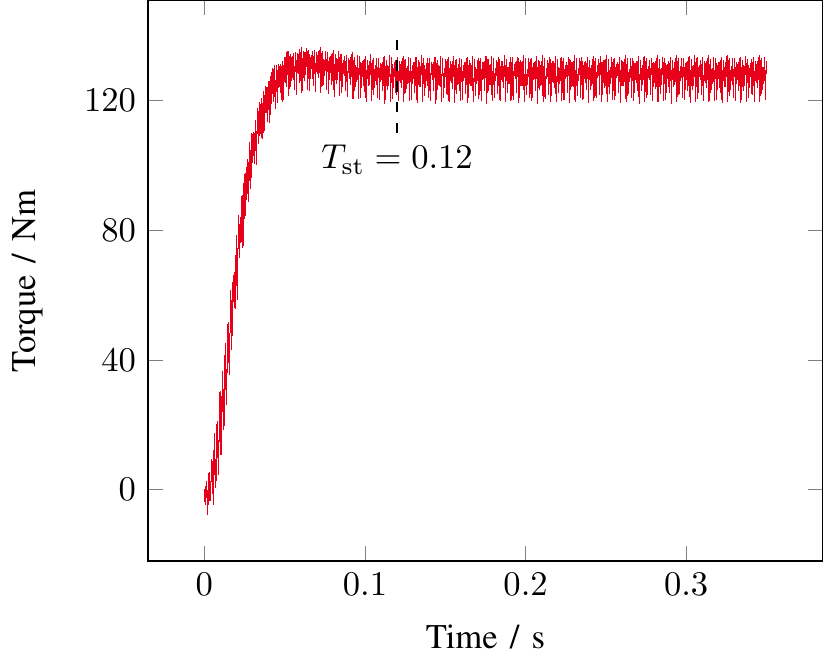}
			\caption{Operating point OP2\label{fig:OP2}}
		\end{subfigure}
		\caption{Torque evolution, calculated with the classical time stepping. $T_\mathrm{st}$ denotes the instant when the steady state is reached.}
		\label{fig:data_OP4_OP1}
	\end{figure*}

To demonstrate the capability of PP-IC we consider two representative operating points of the machine. OP1 describes the operation limit at maximum speed of $18\,000$ rpm while OP2 corresponds to the maximum acceleration at low speed. The torque evolutions, {obtained with the classical time stepping,} are depicted in Figure~\ref{fig:data_OP4_OP1}. We note that within the performed simulations the rotational speed was prescribed by the operating points, which omits solution of mechanical equations \eqref{eq:motion}. %

{Within the time-marching procedure} we consider the steady-state solution to be attained up to a prescribed tolerance {$\varepsilon$} at instant $T_\mathrm{st},$ %
{which belongs to time interval $[t_0+(k^*-1)T,$ $t_0+k^*T].$ %
Period $k^*=\min_{k\geq1} \{k: err(k)\leq\varepsilon\}$ is defined here based on the (relative) periodicity error $err,$
given in terms of torque $\mathbf{T}(t)$ by
\begin{equation}
			err(k)=\frac{|\mathbf{T}(t_0+(k-1)T)-\mathbf{T}(t_0+kT)|}{|\mathbf{T}(t_0+kT)|}.
			\label{eq:err}
\end{equation}
We denote the periodicity error of the calculated torque in the steady state by $err^*=err(k^*).$ 
Performance of the considered acceleration methods will be estimated in comparison to the sequential 
time stepping until $T_\mathrm{st}.$
}

	\subsection{Operating point 1}
We consider the operating point OP1 at the revolution speed of $18\,000$ rpm and the sinusoidal three-phase input current with the amplitude of $125$ A. The resulting mean torque in the steady state has to be determined by the transient eddy current simulation.
	
		First, we apply PP-IC on the time interval {$[0, 0.0311]$ s, i.e., on one electric period in rotor,} using $80$ parallel processors.  
$12$ iterations of the algorithm with the fine step $\delta t=4.629630\cdot10^{-6}$ s produced a periodic solution up to tolerance $\varepsilon=1.6\cdot 10^{-2}$ with respect to torque. The simulation required $1\,968$ effective time steps {to calculate the torque, periodic up to tolerance $\varepsilon=1.6\cdot 10^{-2}$}. {This same value is also a bound for the periodicity error $err^*=1.6\cdot10^{-2}$ of the steady-state solution, obtained from the classical time stepping on period $k^*=9,$ i.e., on $[0.2488, 0.2799]$ s. The effort of the sequential computations is evaluated by $56\,160$ time steps of size $\delta t$, calculated on $[0, 0.26]$ s. Comparing this to the number of effective time steps performed within PP-IC, we get the speedup of factor $28$ due to time-parallelization.} {We would like to note that period $T=0.0311$ s has been empirically determined, more sophisticated approaches could include $T$ as an additional variable to be determined by Parareal.}%

\begin{figure*}[t]
		\centering
		\begin{subfigure}{0.33\linewidth}
			\centering
			\includegraphics[width=.9\linewidth]{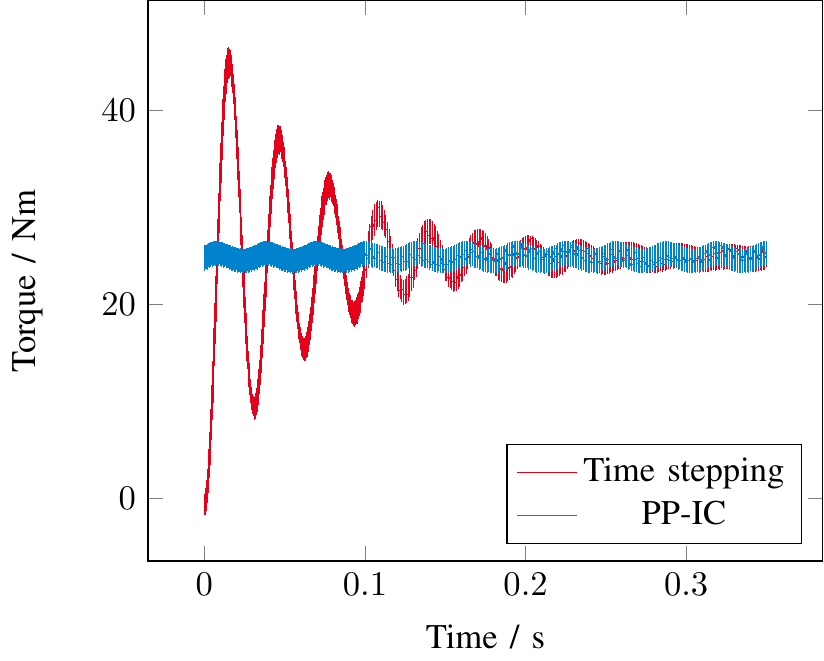}
			\caption{PP-IC for OP1\label{fig:OP1_PPIC}}
		\end{subfigure}%
		\begin{subfigure}{0.33\linewidth}
			\includegraphics[width=0.9\linewidth]{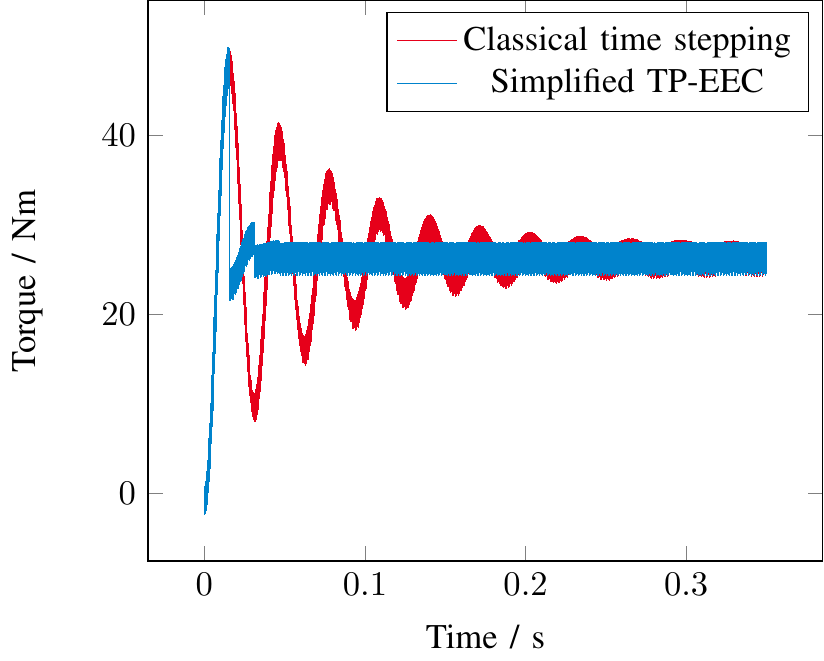}
			\caption{TP-EEC for OP1\label{fig:OP1_EEC}}
		\end{subfigure}%
		\begin{subfigure}{0.33\linewidth}
			\includegraphics[width=0.9\linewidth]{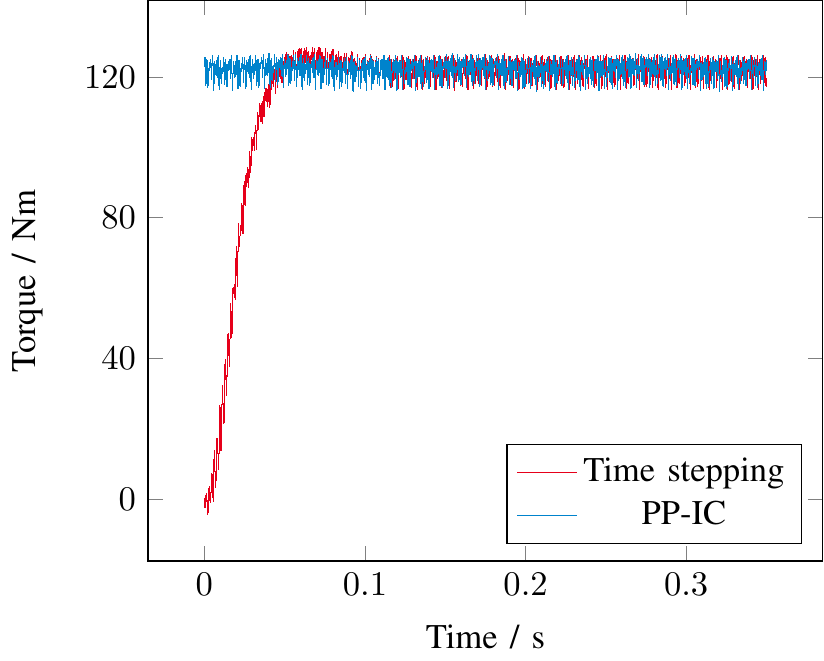}
			\caption{PP-IC for OP2\label{fig:OP2_PPIC}}
		\end{subfigure}%
		\caption{Torque evolution for different methods and operating points in comparison to classical time stepping. 
		{The PP-IC solutions, obtained on one rotor period, are replicated over the whole time interval for visualization purposes.}}
		\label{fig:data_PPIC_EEC}
	\end{figure*}

	The periodic {torque,} obtained with the PP-IC iteration is depicted in Figure~\ref{fig:OP1_PPIC}, {together with the reference time-stepping behavior.} We note that 
	the calculated PP-IC solution {still contains a residual oscillation remains,} {although being almost perfectly periodic. This is in contrast} to the sequential steady-state solution, which eventually flattens out. The deviation (by $0.5$ Nm), however, does not exceed $2\%$ of the mean torque of $25$ Nm. {Therefore,} we consider {this oscillation} to be acceptable, {since it is in a good agreement with the chosen periodicity tolerance $\varepsilon$. }
		
	On the other hand, refining the fine step size {by the factor of $10,$ i.e., using $\delta t=4.629630\cdot10^{-7}$ s,} let the {residual oscillation} vanish {and delivered the steady-state solution, compatible with the sequential calculation of the same time-stepping precision}. {We conclude} that PP-IC {performs well} but is more sensitive to {the} step size {of the fine propagator} than an ordinary (sequential) time domain simulation. 	
	
	{We {now} compare the performance of PP-IC with the simplified TP-EEC method \cite{Takahashi_2010aa}. The main idea of the approach is a successive reduction of the time-stepping solution at every half a period $t=t_0+T/2$ by the average of the initial value $\mathbf{u}(t_0)$ and the final value $\mathbf{u}(t_0+T/2)$ until the steady state is attained. Application of this method to OP1 shows that time stepping with the TP-EEC correction at every $T/2=0.01555$ s converges to the steady state about $7$ times faster than the ordinary time-marching procedure of $56\,160$ steps.} {Time-stepping calculation for OP1, corrected with the simplified TP-EEC, is depicted in Figure~\ref{fig:OP1_EEC}, where the steady-state solution, {periodic up to tolerance $1.2\cdot 10^{-1}$,} is obtained already after the second correction.}

{Please note, TP-EEC is not a parallel method and requires less overall computing power than Parareal.
Furthermore, the speedup provided by TP-EEC could be increased by its joint application with the Parareal algorithm, e.g., via application of Parareal on half-period $[t_0,t_0+T/2]$ followed by the TP-EEC correction at $t=t_0+T/2.$}

	\subsection{Operating point 2}
At OP2 the rotor speed is $1\,000$ rpm and peak value of the phase currents is $160$ A. In contrast to the previously considered example, OP2 has a different transient behavior: there is no significant overshoot of the mean torque (Figure~\ref{fig:OP2}). PP-IC, applied on {one rotor period $[0,0.114]$ s} with $154$ cores and fine step size $\delta t=4.629630\cdot10^{-6}$ s converged in $4$ iterations, reaching the steady-state solution, {periodic up to tolerance $\varepsilon=2\cdot 10^{-3}$ in terms of torque, and thereby requiring calculation of only $1\,256$ effective time steps.}
{On the other hand, classical time stepping led to the torque with relative periodicity error $err^*<10^{-2}$ on period $k^*=2$. Compared to $25\,920$ sequential steps of the same (fine step) size $\delta t,$ performed on $[0,0.12]$ s, a speedup of factor $20$ is achieved with PP-IC.}			
	
	{An attempt to apply the simplified TP-EEC to OP2 unfortunately did not converge. Correction at every half of the period $T/2=0.057$ s set the transient solution even further apart from reaching the steady state. } {We summarize the computational costs of PP-IC and simplified TP-EEC for the two considered operating points, in contrast to the standard sequential time stepping until the steady state, in Table \ref{tab:Comparison}.	}
		\begin{table}[htb]
		\centering
		\begin{tabular}{@{}lccc@{}}
			\toprule
			& Sequential & PP-IC & simplified TP-EEC \\
			\midrule
			OP1 & $56\,160$ & $1\,968$ & $7\,919$\\
			OP2 & $25\,920$ & $1\,256$ & not applicable\\
			\midrule
			\bottomrule
		\end{tabular}		
		\caption{Number of the solved linear systems within PP-IC and simplified TP-EEC, in contrast to the sequential time stepping.} 
		\label{tab:Comparison}
	\end{table}

	\section{Conclusion}\label{sec:conclusion}
	A periodic Parareal algorithm with initial-value coarse problem (PP-IC) is a suitable approach for accelerated attainment of the steady state, since it is applied to a time-periodic problem directly, in contrast to the standard Parareal method, which solves an IVP without taking into account the periodicity condition. Application of PP-IC to an industry-relevant problem illustrated a significant speedup in terms of effective time steps. Numerical experiments illustrated that convergence of PP-IC is more sensitive to the fine step size, compared to the classical time stepping. Nonetheless, the algorithm could be successfully applied to all tested operating points, whereas the simplified TP-EEC method delivered feasible results only in some cases. In the applicable case (OP1) the speedup obtained with the simplified TP-EEC is smaller than that of PP-IC, however the Parareal-based approach needs more (parallel) computing power. {Note that for steady-state calculation of asynchronous machines with methods, based on the periodicity condition, additional research should be dedicated to a suitable choice of the common period $T$.} Further development of efficient numerical methods for accelerated steady-state analysis could be based, e.g., on another parallel-in-time algorithm, tailored to periodic problems, called \textit{periodic Parareal algorithm with periodic coarse problem (PP-PC)} \cite{Gander_2013ab}, where the periodicity constraint on the coarse grid is imposed explicitly.
	
\bibliographystyle{IEEEtran}
\input{references}

\end{document}